\documentclass[conference,10pt,twocolumn]{IEEEtran}
\IEEEoverridecommandlockouts
\usepackage{cite}
\usepackage{amsmath,amssymb,amsfonts,amsthm}
\usepackage{graphicx}
\usepackage{textcomp}
\usepackage{xcolor}
\usepackage{dsfont}
\usepackage{bm}
\usepackage{acronym}
\usepackage[left=0.625in, right=0.625in, top=0.7in, bottom=1.0in]{geometry}

\usepackage{tabu}
\usepackage{tikz}
\usepackage{lipsum}

\newtheorem{theorem}{Theorem}

\newtheorem{definition}{Definition}

\newtheorem{remark}{Remark}

\usepackage{algorithm}
\usepackage{algpseudocode}

\title{Over-the-Air Collaborative Inference with Feature Differential Privacy}

\author{\IEEEauthorblockN{Mohamed Seif \quad Yuqi Nie \quad Andrea J. Goldsmith \quad H. Vincent Poor\\
\thanks{This work has been supported in part by U.S National Science Foundation under Grants CNS-2147631, ECCS-2335876, a grant from Princeton Language \& Intelligence, and the AFOSR Grant Award no. 002484665}
}
\IEEEauthorblockA{Department of Electrical and Computer Engineering, \\ 
Princeton University, Princeton, NJ, 08544\\
Email: $\{\textit{mseif, ynie, goldsmith, poor}\}$@princeton.edu}
}

\begin{document}

\maketitle

\begin{abstract}
 Collaborative inference in next-generation networks can enhance Artificial Intelligence (AI) applications, including autonomous driving, personal identification, and activity classification. This method involves a three-stage process: a) data acquisition through sensing, b) feature extraction, and c) feature encoding for transmission. Transmission of the extracted features entails the potential risk of exposing sensitive personal data. To address this issue, in this work a new privacy-protecting collaborative inference mechanism is developed. Under this mechanism, each edge device in the network protects the privacy of extracted features before transmitting them to a central server for inference. This mechanism aims to achieve two main objectives while ensuring effective inference performance: 1) reducing communication overhead, and 2) maintaining strict privacy guarantees during features transmission.
\end{abstract}


\section{Introduction}

Artificial intelligence (AI) is expected to be a key enabler for new applications in next-generation networks \cite{saad2019vision, letaief2019roadmap, kairouz2021advances}. For example, it can enable low-latency inference and sensing applications, including autonomous driving, personal identification, and activity classification (to name a few). Two conventional AI paradigms are commonly used in practice for these applications: 1) On-device inference that locally performs AI-based inference. This approach suffers from high computation overhead relative to device capabilities, and 2) On-server inference, where edge devices upload their raw data to a central server to perform a global inference task. The latter approach may compromise the data privacy of individuals, and it suffers from high communication overhead. To remedy these challenges, edge-device collaborative inference is a compelling solution. In this setting,  joint inference is divided into three modules: a) sensing for data acquisition, b) feature extraction, and 3) feature encoding for transmission. Leakage of fine-grained information about individuals is a risk that must be considered while designing such task-oriented communication systems.

To address this challenge, we develop a new private collaborative inference mechanism wherein each edge device in the network protects the sensitive information of  extracted features before transmission to a central server for inference. The key design objectives of this approach are two-fold: 1) minimizing the communication overhead and 2) maintaining rigorous privacy guarantees for transmission of features over a communication network, while providing satisfactory inference performance. Our wireless distributed machine learning transmission scheme, inspired by the findings in \cite{seif2020wireless}, optimizes bandwidth, computational efficiency, and differential privacy (DP) by leveraging the superposition nature of the wireless channel. This approach, in contrast to tradition orthogonal signaling methods, offers enhanced privacy and expedited task accuracy. Further strengthening our scheme, we incorporate additional novel strategies involving aggregated perturbation coupled with device sampling. This method introduces controlled noise to the aggregated data from multiple devices. The combination of wireless superposition, edge device sampling, and aggregated perturbation forms a comprehensive and efficient transmission framework for the wireless collaborative inference problem (see recent works in the literature \cite{yilmaz2022over, liu2023over, chen2023view}).








\begin{figure*}[t]
    \centering
    \includegraphics[width= 1.6\columnwidth]{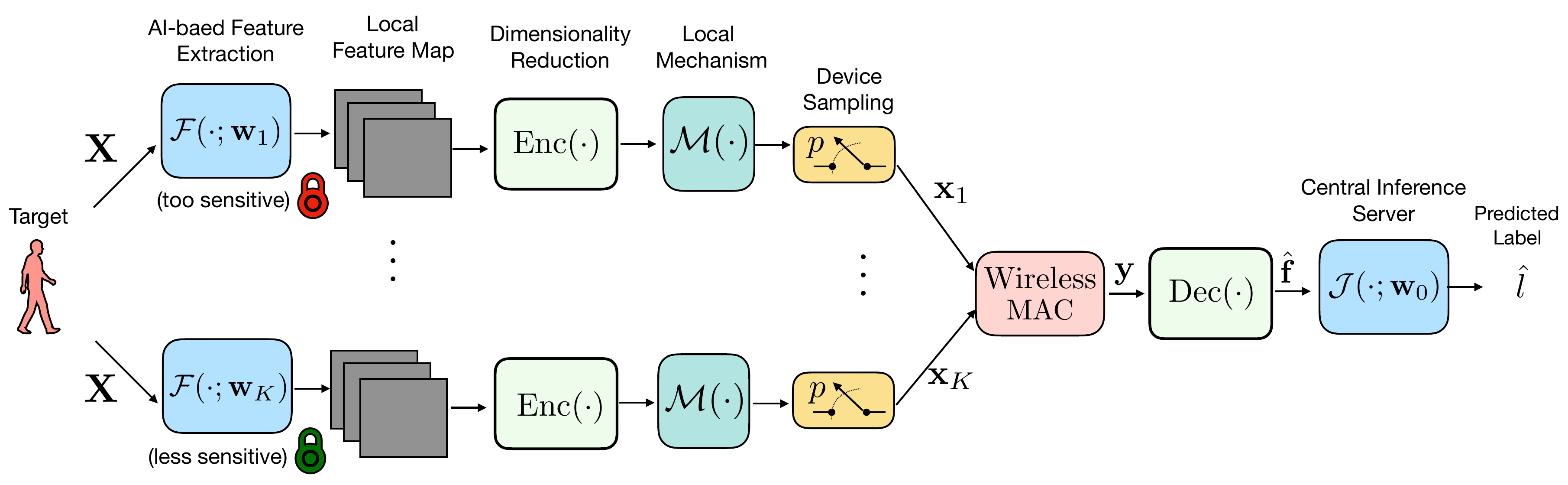}
    \caption{\small{Illustration of the private task-inference framework: Each edge device extracts features from the observed input that preserves some relevant information for classification while satisfies rigorous feature DP levels. Then, each device forwards the processed features over a communication channel to be processed by the central inference server.}}
    \label{fig:proposed_inference_model}
\end{figure*}

\section{System Model \& Problem Statement}

\subsection{Communication Channel Model}

We consider a single-antenna distributed inference system with $K$ edge devices and a central inference server. The edge devices are connected to the inference server through a wireless multiple-access channel with fading on each link. Let $\mathcal{K}$ be a random subset of edge devices that transmit to the server. The input-output relationship can be expressed as
\begin{align}
    \mathbf{y} = \sum_{k \in \mathcal{K}} h_{k} \mathbf{x}_{k} + \mathbf{m},
\end{align}
where $\mathbf{x}_{k} \in \mathds{R}^{d}$ is the transmitted signal by device $k$, $\mathbf{y}$ is the received signal at the edge server, and $h_{k} \geq 0$ is the  channel coefficient between the $k$th device and the server. We assume a block flat-fading channel, where
the channel coefficient remains constant within the duration of a communication block. We denote $\mathbf{m} \in \mathds{R}^{d}$ as the receiver noise whose elements are independent and identically distributed (i.i.d.) according to Gaussian distribution with zero-mean and variance $\sigma_{m}^{2}$.



\subsection{Distributed Inference Model}

A pre-trained sub-model is deployed on each edge device $k$ that takes the captured image as input and outputs a feature map (or tensor) of real-valued features. Denote the vectorized version of the tensor as $\mathbf{f}_{k} \in \mathds R^{d}$. The edge server performs a multi-view \textit{average} pooling operation on the received local feature maps $\mathbf{f}_{k}$'s to obtain a \textit{global} feature map ${\mathbf{f}}^{*}$ and feeds it to the pre-trained server model to perform a classification task. The average pooled feature is obtained as 
\begin{align}
    \mathbf{f}^{*} & = \frac{1}{K} \sum_{k=1}^{K} \mathbf{f}_{k}.
\end{align}

\section{Main Reuslts \& Discussions} \label{sec:transmission_scheme}

In this section, we first introduce our proposed transmission scheme.  We then outline the scheme's privacy guarantees as described in Theorem \ref{thm:feature_privacy_guarantee}. Finally, we establish a lower bound for the classification accuracy of our approach in Theorem \ref{thm:lower_bound_classification}. We summarize the scheme in Algorithms $1$ and $2$.

\subsection{Proposed Transmission Scheme}



\noindent $(1)$ \textbf{Feature extraction and dimensionality reduction.} Each device $k$ first performs feature extraction\footnote{In this paper, we neglect the data aquisition error.} to obtain an informative representation of the common target  $\mathbf{X}$. This is followed by dimensionality reduction, executed via an encoding operation. The dimensionality reduction process can be represented as
\begin{align}
\mathbf{z}_{k} = \mathbf{W}_{k} \mathbf{f}_{k}(\mathbf{X}),
\end{align}
where $\mathbf{W}_{k} \in \mathds{R}^{r \times d}$ denotes the weight matrix of the encoder for device $k$ where $r \leq d$, and $\mathbf{f}_{k}(\mathbf{X}) \in \mathds{R}^{d}$. \\


\noindent $(2)$ \textbf{Local perturbation noise for privacy.}  Each device $k$ computes a noisy version of its extracted feature as 
\begin{align}
    \tilde{\mathbf{z}}_{k} = w_{k} \mathbf{z}_{k} + \mathbf{n}_{k}, \nonumber 
\end{align}
where $\mathbf{n}_{k} \sim \mathcal{N}(0, \sigma_{k} \mathbf{I}_{q})$ is the artificial noise for privacy. We further assume that the norm of feature vector is bounded by some constant $C_{k} \geq 0$, and in order to ensure that we normalize the feature vector by $C_{k}$, i.e., $\mathbf{z}_{k} := \min \left(1, C_{k}/\| \mathbf{z}_{k} \|_{2}\right)  \cdot \mathbf{z}_{k}$. Finally, $w_{k} \geq 0$ is a weight coefficient of the $k$th device. \\

\noindent $(3)$ \textbf{Pre-processing for transmission.}  The transmitted signal of device $k$ is given as:
\begin{align}
    \mathbf{x}_{k} =  \begin{cases}  \frac{\alpha_{k} }{p_{k}}  \tilde{\mathbf{z}}_{k}, & \text{w.p.}~p_{k}\\
    \mathbf{0}, & \text{otherwise},
    \end{cases}
    \label{eq:inputsignal}
\end{align}
where $\alpha_{k}$ is a scaling factor. If a device is not participating, it does not transmit anything. Note that we multiply the transmitted signal by $1/p_{k}$ to ensure that the estimated signal (i.e., feature map) seen at the server is unbiased. \\








\noindent $(4)$ \textbf{Features aggregation at the edge server.} The received signal at the inference server is given as:
\begin{align}
    \mathbf{y} 
    = \sum_{k \in \mathcal{K}}  \frac{h_{k} \alpha_{k}w_{k}}{p_{k}}  \mathbf{z}_{k}  +  \sum_{k \in \mathcal{K}} \frac{h_{k} \alpha_{k}}{p_{k}} \mathbf{n}_{k} +  \mathbf{m}.\label{eq:output}
\end{align}
All edge devices pick the coefficients $\alpha_{k}$'s to align their transmitted local features. Specifically, each device $k$ picks $\alpha_{k}$ so that ${h_{k} \alpha_{k}}/{p_{k}} = \gamma, \forall k \in \mathcal{K}$, where $\gamma$ represents the chosen alignment constant. \\



\noindent $(5)$ \textbf{Post-processing at the edge server.}  Subsequently, the server performs the following sequence of post-processing:
\begin{align}
    \hat{\mathbf{z}} = \frac{1}{\gamma} \mathbf{y} = \sum_{k \in \mathcal{K}} w_{k} \mathbf{z}_{k} + \sum_{k \in \mathcal{K}} n_{k} + \frac{1}{\gamma}  \mathbf{m}. \label{eqn:post_processing}
\end{align}

\noindent $(6)$ \textbf{Decode the aggregated signal.} The server then decodes the post-processed signal $\hat{\mathbf{z}}$ as follows:
\begin{align}
    \hat{\mathbf{f}} & = \mathbf{D}  \hat{\mathbf{z}} = \mathbf{D} \sum_{k \in \mathcal{K}} w_{k}  \mathbf{W_{k}} \mathbf{f}_{k} + \mathbf{D} \sum_{k \in \mathcal{K}} n_{k} + \frac{1}{\gamma} \mathbf{D} \mathbf{m},
\end{align}
where $\mathbf{D} \in \mathds{R}^{d \times r}$ is the decoding matrix deployed at the central server.

\begin{algorithm}[t]
\caption{Differentially Private Feature Extraction}
\label{alg:private_extraction}
\begin{algorithmic}[1]
    \State \textbf{Input:} Collect observations $\{\mathbf{X}_{k}\}_{k=1}^{K}$ of the target $\mathbf{X}$ 
    \For{each edge device $k \in \mathcal{K} $ in parallel}
        \State Perform feature extraction on the observed target using the pre-trained model $\mathbf{w}_{k}$: $\mathbf{f}_{k} = \mathcal{F}(\mathbf{X}_{k}; \mathbf{w}_{k})$
        \State Perform dimensionality reduction: $\mathbf{z}_{k} = \mathbf{W}_{k} \mathbf{f}_{k}$
        \State Clip the feature vector: $\mathbf{z}_{k} \gets \min \left(1, \frac{C_{k}}{\| \mathbf{z}_{k} \|_{2}}\right)  \cdot \mathbf{z}_{k}$
        \State Perturb the feature vector via Gaussian mechanism: $\tilde{\mathbf{z}}_{k} \gets w_{k} \mathbf{z}_{k} + \mathbf{n}_{k}$, where $w_{k}$ is a weight coefficient 
    \EndFor
    \State \textbf{Output:} $\tilde{\mathbf{z}}_{1}, \tilde{\mathbf{z}}_{2}, \ldots, \tilde{\mathbf{z}}_{K}$
\end{algorithmic}
\end{algorithm}

\begin{algorithm}[t]
\caption{Features Aggregation and Model Inference}
\label{alg:model_inference}
\begin{algorithmic}[1]
    \State \textbf{Input:}  $\tilde{\mathbf{z}}_{1}, \tilde{\mathbf{z}}_{2}, \ldots, \tilde{\mathbf{z}}_{K}$
        \State The server performs pooling on operation on the received features to obtain a global feature map according to \eqref{eqn:post_processing}
        \State Decode the global feature map: $\hat{\mathbf{f}} = \mathbf{D}  \hat{\mathbf{z}}$
        \State The global feature map is then fed in the ML model at the server for inference: $\hat{l} = \mathcal{J}(\mathbf{w}_{0};\hat{\mathbf{f}})$
        \State \textbf{Output:} Predicted label $\hat{l}$
\end{algorithmic}
\end{algorithm}

\subsection{Feature differential privacy analysis}

We analyze the privacy level achieved by our proposed scheme that adds artificial noise perturbations to privatize its local data. More precisely, we analyze the privacy leakage under an additive noise mechanism drawn from a Gaussian distribution \cite{dwork2014algorithmic}. We next describe the thereat model. \\

\noindent \textbf{Privacy Threat Model}: In the collaborative inference framework, we assume that the central inference server is \textit{honest but curious}. It is honest because it follows the procedure accordingly, but it might learn sensitive information about features. The inference results are released to potentially untrustworthy third parties, heightening privacy concerns. Our focus is on ensuring differential privacy (DP). DP maintains that algorithm outputs (i.e., the task predictions) are indistinguishable when inputs (i.e., the features) differ slightly. Formally, the \textit{feature} DP guarantee can be described as follows: 

\begin{definition} 
[$(\epsilon, \delta)$-feature DP] Let $\mathcal{D} \triangleq \mathcal{F}_{1} \times \mathcal{F}_{2}  \times \cdots \times \mathcal{F}_{K} $ be the collection of all possible features of a common object $\mathbf{X}$.  A randomized mechanism $\mathcal{M}: \mathcal{D} \rightarrow \mathds{R}^{d}$ is $(\epsilon, \delta)$-feature DP if for any two neighboring $D, D' \in \mathcal{F}$, and any measurable subset $\mathcal{S} \subseteq \text{Range}(\mathcal{M})$, we have
\begin{align}
    \operatorname{Pr}(\mathcal{M}(D) \in \mathcal{S}) \leq e^{\epsilon} \operatorname{Pr}(\mathcal{M}(D') \in \mathcal{S}) + \delta.
\end{align}
Here, we refer a pair of neighboring datasets $D, D' \in \mathcal{D}$ if $D'$ can be obtained from $D$ by removing one element, i.e., the feature extracted by the $k$th device. The setting when $\delta = 0$ is referred as pure $\epsilon$-feature DP. 
\end{definition}



\begin{theorem}\label{thm:feature_privacy_guarantee} (Privacy Guarantee) For each edge device $k$ participates with probability $p_{k} \geq 0$ and utilizes local mechanism with an importance weight $w_{k} \geq 0$.  The privacy guarantee for the $k$th feature is given as
\begin{align}
    \epsilon_{k} \leq \log \left[ 1 + \frac{p_{k}}{1-\delta'} \left( e^{\frac{c_{k}}{\sqrt{\bar{\mu} - t}}} -1 \right)  \right],  \tilde{\delta}_{k} = \delta' + \frac{p_{k} \delta}{1 - \delta'},
\end{align}
for any $\delta, \delta' \in (0,1]$ such that $\operatorname{Pr}(|\mu - \bar{\mu}| \geq t) \leq \delta'$  where $\mu \triangleq \sum_{i = 1}^{K} \tau_{i} \sigma_{i}^{2}$, $\tau_{i} \sim \operatorname{Bern}(p_{i})$, $\bar{\mu} \triangleq \sum_{i=1}^{K} p_{i} \sigma_{i}^{2}$, and $c_{k} \triangleq \gamma w_{k} C_{k} \sqrt{2 \log(1.25/\delta)}$. Further, for a given $\delta'$, we choose the parameter $t$ as $    t = \frac{\max_{k} \sigma_{k}^{2}}{{2/\log(2/\delta')}} + \frac{\sqrt{{\max_{k} \sigma_{k}^{2}}/{9}  + 4 (\sum_{k \in [n]} p_{k} (1-p_{k}) \sigma_{k}^{4})/\log(2/\delta')}}{2/\log(2/\delta')}$.
\end{theorem}

 \begin{remark}
Central to our approach is the custom adaptation of privacy guarantees to the feature's varying sensitivity levels. To address the diversity in data sensitivity and privacy needs, we introduce a system of weight coefficients \(w_{k}\) and clipping threshold $C_{k}$ for each feature vector, reflecting their respective DP sensitivities. This enables a tailored privacy protection approach. The development of a device-specific DP leakage metric, \(\epsilon_{k}\), incorporates these customized parameters, allowing for privacy adjustments that align with the distinct sensitivities of the devices' contributed feature vectors.
 \end{remark}





\subsection{Classification Accuracy}

The goal is to analyze the inter-relationship between accuracy and aggregation error due to the randomness of the privacy-preserving perturbation mechanism and sampling procedure. The Mean Squared Error (MSE) can be readily obtained as follows:
\begin{align}
& \operatorname{MSE}   \triangleq \mathds{E} \left[ \|\hat{\mathbf{f}} - \mathbf{f}^{*}\|_{2}^{2} \right] \nonumber \\ 
& \leq d \cdot \|\mathbf{D}\|_{F}^{2} \cdot  \left[ \sum_{k=1}^{K} \frac{2 p_{k} w_{k}^{2} C_{k}^{2}}{\epsilon_{k}^{2}} \log\left(\frac{1.25}{\delta_{k}}\right)  +  \frac{\sigma_{m}^{2}}{\gamma^{2}} \right] \nonumber \\
& +  \sum_{k = 1}^{K} \bigg[ (w_{k}^{2} p_{k} - 2 w_{k} p_{k} + 1) \|\mathbf{D}\|_{F}^{2} \|\mathbf{W}_{k}\|_{F}^{2} \|\mathbf{f}_{k}\|_{2}^{2} \bigg] \nonumber \\
&   +  \sum_{k < j} \bigg[ (p_{k} p_{j} w_{k} w_{j} - p_{k} w_{k} - p_{j} w_{j} + 1) \mathbf{f}_{k}^{T} \mathbf{W}_{k}^{T} \mathbf{D}^{T} \mathbf{D} \mathbf{W}_{j} \mathbf{f}_{j} \bigg], \nonumber 
\end{align}
where the first term in the MSE expression represents the effective noise seen at the inference server, which includes contributions from both channel noise and local perturbation noise introduced for privacy. The second and third terms quantify the approximation error resulting from the application of weight coefficients and the stochastic nature of device participation. It is crucial to highlight that the expectation in the MSE calculation accounts for the randomness introduced by both the variable participation of devices and the variations in local perturbation noise and channel noises. It is worth highlighting that the third term captures the correlations between features $\mathbf{f}_{k}$'s since they are extracted from the same target $\mathbf{X}$.




\begin{remark}
Note that the deployed central server model has an intrinsic classification margin $\Delta$ the feature space, that is defined as the minimum distance in which the model classifies correctly the pooled feature $\hat{\mathbf{f}}$ when  $\mathds{E} \left[ \|\hat{\mathbf{f}} - \mathbf{f}^{*}\|_{2}^{2} \right]  \leq \Delta$.
\end{remark}

We next establish a lower bound for the classification accuracy of our proposed scheme. This approach is grounded in the concept of the \textit{classification margin}, as detailed in \cite{sokolic2017robust}.

\begin{theorem} [Classification Accuracy] \label{thm:lower_bound_classification} The lower bound on the classification accuracy for our proposed privacy-preserving method can be expressed as
\begin{align}
P(\hat{l} = l ) \geq \max \left[0, P_{0} \cdot \left(1 - \left(\frac{\operatorname{MSE}}{\Delta}\right)^{2} \right)\right],
\end{align}
where $P_{0}$ represents the classification accuracy of the local model, and $\Delta$ represents the inherent classification margin.
\end{theorem}



\begin{figure}[t]
	\centering
    {\includegraphics[width=0.75\columnwidth]{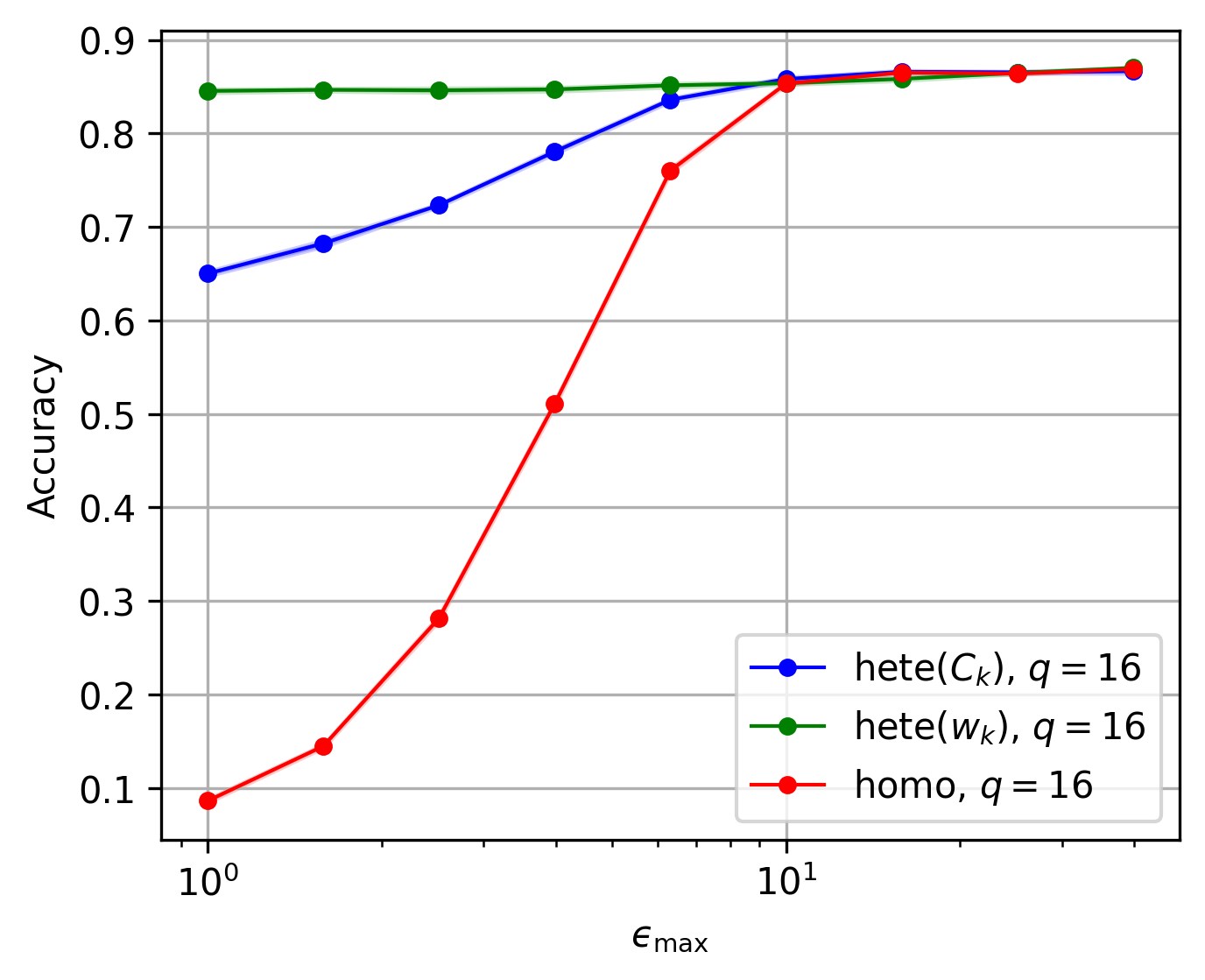}}
    \caption{\small{Impact of customizing privacy levels on the classification accuracy for $r = q \times 7 \times 7$, where $q = 16$.}}
    \label{fig:impact_of_hete_privacy_small}
    \vspace{-10pt}
\end{figure}

\begin{figure}[t]
	\centering
    {\includegraphics[width=0.75\columnwidth]{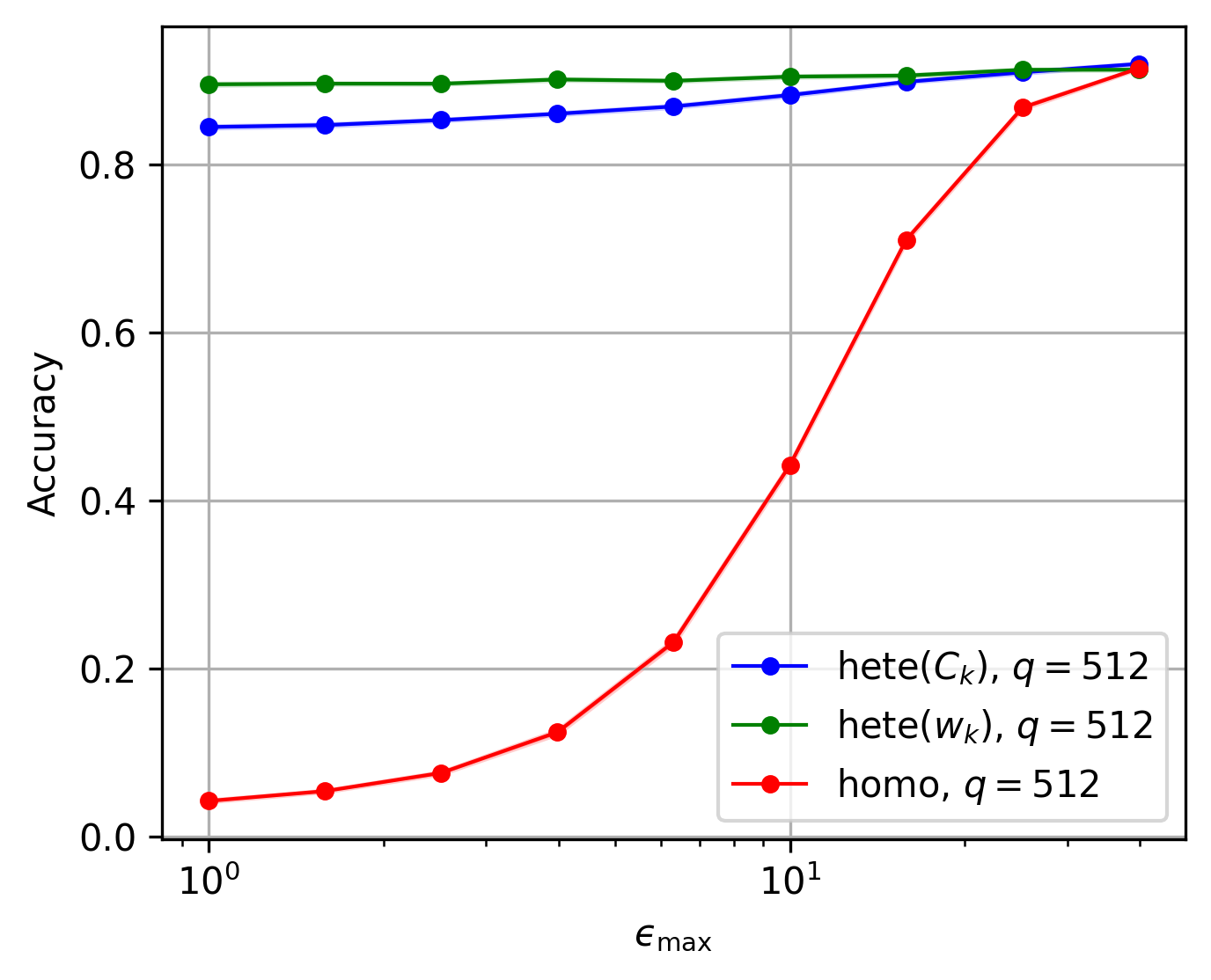}}
    \caption{\small{Impact of customizing privacy levels on the classification accuracy for $r = q \times 7 \times 7$, where $q = 512$.}}
    \label{fig:impact_of_hete_privacy_large}
    \vspace{-10pt}
\end{figure}








     










\section{Experiments and Performance Analysis}



In this section, we conduct experiments to assess the performance of the proposed private collaborative inference scheme. We adopt a Rician channel model with a variance of $\sigma^{2}_{m} = 0.1$ to simulate fading channels \cite{goldsmith2005wireless}. Our setup includes $K = 12$ devices, each with a default transmit power of $P_{k} = 30$ dBm, a connection probability of $p_k = 0.9$, and an equal weight of $w_{k} = 1/K = 1/12$. The perturbation noise level is set at $\sigma^{2}_{k}=0.1$, with privacy parameters $\delta = 10^{-5}$ and $\delta'= 10^{-5}$, and an alignment constant of $\gamma=1$. Additionally, we employ feature clipping on each device before transmitting the local feature to a central server, with a clipping threshold of $C_k = 10^2$.

The encoding and decoding matrices are implemented via one hidden layer neural networks for two different dimensions, $16 \times 7 \times 7$ and $512 \times 7 \times 7$, to facilitate the proposed private collaborative inference scheme\footnote{It is worth highlighting that performance worsens for higher dimensions due to the increase in perturbation noise.}. We develop a Multi-View Convolutional Neural Network (MVCNN) architecture utilizing the ModelNet dataset, known for its multi-view images of objects such as sofas and tables, and integrating the VGG11 model. In our design, the VGG11 model is partitioned prior to the linear classifier stage, positioning the classifier at the server for ultimate decision-making and distributing the remaining VGG11 components across sensor nodes for feature extraction, optimized for average pooling. Our study concentrates on a ModelNet subset encompassing $40$ object classes, captured using an array of $12$ cameras arranged to provide a $30$-degree separation between adjacent sensors for thorough and diverse object perspectives. Each sensor, equipped with the adapted VGG11 model, produces feature maps as tensors of dominions $16 \times 7 \times 7$ and $512 \times 7 \times 7$, respectively.

In Fig. \ref{fig:impact_of_hete_privacy_small} and \ref{fig:impact_of_hete_privacy_large}, we reveal the critical need for customizing privacy levels to optimize performance, acknowledging that not all transmitted features bear the same sensitivity. With half of the edge devices processing data with sensitive attributes and the other half not, we highlight two strategic avenues for enhancement: \textit{optimizing} the weight coefficients \(w_{k}\)'s, or \textit{refining} the clipping parameters \(C_{k}\). Upon comparing these approaches with uniform privacy models, where \(w_{k} = 1/K\) and \(C_{k} = C\), our methods demonstrate superior performance in scenarios demanding stringent privacy. It is noteworthy that uniform privacy approaches only approximate the efficacy of our tailored strategies in the low-privacy regime (i.e., large $\epsilon$). This insight aligns with findings from \cite{shlezinger2022collaborative}, which critique the limitations of incorporating DP in inference systems where data privacy prevails, thus highlighting that traditional DP mechanisms may not always ensure optimal utility.

\section{Conclusions}

In this paper, we investigate the problem of collaborative inference over wireless channels. We demonstrate the synergistic benefits of edge device sampling and wireless aggregation on the privacy guarantees of feature transmissions. We also provide a lower bound on the classification accuracy as a function of the channel parameters, privacy level, and feature dimensions. Our experiments on a real-world dataset validate the efficacy of our proposed transmission scheme.

\bibliographystyle{IEEEtran}
\bibliography{myreferences}

\end{document}